
%
%
\def\unredoffs{} \def\redoffs{\voffset=-.31truein\hoffset=-.48truein}
\def\speclscape{}
%
%
%
%
%
\newbox\leftpage \newdimen\fullhsize \newdimen\hstitle \newdimen\hsbody
\tolerance=1000\hfuzz=2pt
\catcode`\@=11 
\def\bigans{b }
\def\answ{b }
%
\ifx\answ\bigans\message{(This will come out unreduced.}
\magnification=1200\unredoffs\baselineskip=16pt plus 2pt minus 1pt
\hsbody=\hsize \hstitle=\hsize 
\else\message{(This will be reduced.} \let\l@r=L
\magnification=1000\baselineskip=16pt plus 2pt minus 1pt \vsize=7truein
\redoffs \hstitle=8truein\hsbody=4.75truein\fullhsize=10truein\hsize=\hsbody
\output={\ifnum\pageno=0 
  \shipout\vbox{\speclscape{\hsize\fullhsize\makeheadline}
    \hbox to \fullhsize{\hfill\pagebody\hfill}}\advancepageno
  \else
  \almostshipout{\leftline{\vbox{\pagebody\makefootline}}}\advancepageno
  \fi}
\def\almostshipout#1{\if L\l@r \count1=1 \message{[\the\count0.\the\count1]}
      \global\setbox\leftpage=#1 \global\let\l@r=R
 \else \count1=2
  \shipout\vbox{\speclscape{\hsize\fullhsize\makeheadline}
      \hbox to\fullhsize{\box\leftpage\hfil#1}}  \global\let\l@r=L\fi}
\fi
%
\newcount\yearltd\yearltd=\year\advance\yearltd by -1900

\def\Title#1#2{\nopagenumbers\abstractfont\hsize=\hstitle\rightline{#1}%
\vskip 1in\centerline{\titlefont #2}\abstractfont\vskip .5in\pageno=0}
\def\Date#1{\vfill\leftline{#1}\tenpoint\supereject\global\hsize=\hsbody%
\footline={\hss\tenrm\folio\hss}}
%

\def\draftmode{\message{ DRAFTMODE }\def\draftdate{{\rm preliminary draft:
\number\month/\number\day/\number\yearltd\ \ \hourmin}}%
\headline={\hfil\draftdate}\writelabels\baselineskip=20pt plus 2pt minus 2pt
 {\count255=\time\divide\count255 by 60 \xdef\hourmin{\number\count255}
  \multiply\count255 by-60\advance\count255 by\time
  \xdef\hourmin{\hourmin:\ifnum\count255<10 0\fi\the\count255}}}
\def\nolabels{\def\wrlabeL##1{}\def\eqlabeL##1{}\def\reflabeL##1{}}
\def\writelabels{\def\wrlabeL##1{\leavevmode\vadjust{\rlap{\smash%
{\line{{\escapechar=` \hfill\rlap{\sevenrm\hskip.03in\string##1}}}}}}}%
\def\eqlabeL##1{{\escapechar-1\rlap{\sevenrm\hskip.05in\string##1}}}%
\def\reflabeL##1{\noexpand\llap{\noexpand\sevenrm\string\string\string##1}}}
\nolabels
%
\global\newcount\secno \global\secno=0
\global\newcount\meqno \global\meqno=1
\def\newsec#1{\global\advance\secno by1\message{(\the\secno. #1)}
\global\subsecno=0\eqnres@t\noindent{\bf\the\secno. #1}
\writetoca{{\secsym} {#1}}\par\nobreak\medskip\nobreak}
\def\eqnres@t{\xdef\secsym{\the\secno.}\global\meqno=1\bigbreak\bigskip}
\def\sequentialequations{\def\eqnres@t{\bigbreak}}\xdef\secsym{}
\global\newcount\subsecno \global\subsecno=0
\def\subsec#1{\global\advance\subsecno by1\message{(\secsym\the\subsecno. #1)}
\ifnum\lastpenalty>9000\else\bigbreak\fi
\noindent{\it\secsym\the\subsecno. #1}\writetoca{\string\quad
{\secsym\the\subsecno.} {#1}}\par\nobreak\medskip\nobreak}
\def\appendix#1#2{\global\meqno=1\global\subsecno=0\xdef\secsym{\hbox{#1.}}
\bigbreak\bigskip\noindent{\bf Appendix #1. #2}\message{(#1. #2)}
\writetoca{Appendix {#1.} {#2}}\par\nobreak\medskip\nobreak}
%
%
\def\eqnn#1{\xdef #1{(\secsym\the\meqno)}\writedef{#1\leftbracket#1}%
\global\advance\meqno by1\wrlabeL#1}
\def\eqna#1{\xdef #1##1{\hbox{$(\secsym\the\meqno##1)$}}
\writedef{#1\numbersign1\leftbracket#1{\numbersign1}}%
\global\advance\meqno by1\wrlabeL{#1$\{\}$}}
\def\eqn#1#2{\xdef #1{(\secsym\the\meqno)}\writedef{#1\leftbracket#1}%
\global\advance\meqno by1$$#2\eqno#1\eqlabeL#1$$}
%
\newskip\footskip\footskip14pt plus 1pt minus 1pt 
\def\footnotefont{\ninepoint}\def\f@t#1{\footnotefont #1\@foot}
\def\f@@t{\baselineskip\footskip\bgroup\footnotefont\aftergroup\@foot\let\next}
\setbox\strutbox=\hbox{\vrule height9.5pt depth4.5pt width0pt}
\global\newcount\ftno \global\ftno=0
\def\foot{\global\advance\ftno by1\footnote{$^{\the\ftno}$}}
%
\newwrite\ftfile
\def\footend{\def\foot{\global\advance\ftno by1\chardef\wfile=\ftfile
$^{\the\ftno}$\ifnum\ftno=1\immediate\openout\ftfile=foots.tmp\fi%
\immediate\write\ftfile{\noexpand\smallskip%
\noexpand\item{f\the\ftno:\ }\pctsign}\findarg}%
\def\footatend{\vfill\eject\immediate\closeout\ftfile{\parindent=20pt
\centerline{\bf Footnotes}\nobreak\bigskip\input foots.tmp }}}
\def\footatend{}
%
%
\global\newcount\refno \global\refno=1
\newwrite\rfile
\def\ref{[\the\refno]\nref}
\def\nref#1{\xdef#1{[\the\refno]}\writedef{#1\leftbracket#1}%
\ifnum\refno=1\immediate\openout\rfile=refs.tmp\fi
\global\advance\refno by1\chardef\wfile=\rfile\immediate
\write\rfile{\noexpand\item{#1\ }\reflabeL{#1\hskip.31in}\pctsign}\findarg}
\def\findarg#1#{\begingroup\obeylines\newlinechar=`\^^M\pass@rg}
{\obeylines\gdef\pass@rg#1{\writ@line\relax #1^^M\hbox{}^^M}%
\gdef\writ@line#1^^M{\expandafter\toks0\expandafter{\striprel@x #1}%
\edef\next{\the\toks0}\ifx\next\em@rk\let\next=\endgroup\else\ifx\next\empty%
\else\immediate\write\wfile{\the\toks0}\fi\let\next=\writ@line\fi\next\relax}}
\def\striprel@x#1{} \def\em@rk{\hbox{}}
\def\lref{\begingroup\obeylines\lr@f}
\def\lr@f#1#2{\gdef#1{\ref#1{#2}}\endgroup\unskip}

\def\addref#1{\immediate\write\rfile{\noexpand\item{}#1}} 
\def\startrefs#1{\immediate\openout\rfile=refs.tmp\refno=#1}
\def\xref{\expandafter\xr@f}\def\xr@f[#1]{#1}
\def\refs#1{\count255=1[\r@fs #1{\hbox{}}]}
\def\r@fs#1{\ifx\und@fined#1\message{reflabel \string#1 is undefined.}%
\nref#1{need to supply reference \string#1.}\fi%
\vphantom{\hphantom{#1}}\edef\next{#1}\ifx\next\em@rk\def\next{}%
\else\ifx\next#1\ifodd\count255\relax\xref#1\count255=0\fi%
\else#1\count255=1\fi\let\next=\r@fs\fi\next}
%

%
\newwrite\ffile\global\newcount\figno \global\figno=1
\def\fig{fig.~\the\figno\nfig}
\def\nfig#1{\xdef#1{fig.~\the\figno}%
\writedef{#1\leftbracket fig.\noexpand~\the\figno}%
\ifnum\figno=1\immediate\openout\ffile=figs.tmp\fi\chardef\wfile=\ffile%
\immediate\write\ffile{\noexpand\medskip\noexpand\item{Fig.\ \the\figno. }
\reflabeL{#1\hskip.55in}\pctsign}\global\advance\figno by1\findarg}
\def\xfig{\expandafter\xf@g}\def\xf@g fig.\penalty\@M\ {}
\def\figs#1{figs.~\f@gs #1{\hbox{}}}
\def\f@gs#1{\edef\next{#1}\ifx\next\em@rk\def\next{}\else
\ifx\next#1\xfig #1\else#1\fi\let\next=\f@gs\fi\next}
\newwrite\lfile
{\escapechar-1\xdef\pctsign{\string\%}\xdef\leftbracket{\string\{}
\xdef\rightbracket{\string\}}\xdef\numbersign{\string\#}}

\def\writestop{\def\writestoppt{\immediate\write\lfile{\string\pageno%
\the\pageno\string\startrefs\leftbracket\the\refno\rightbracket%
\string\def\string\secsym\leftbracket\secsym\rightbracket%
\string\secno\the\secno\string\meqno\the\meqno}\immediate\closeout\lfile}}
\def\writestoppt{}\def\writedef#1{}
\def\seclab#1{\xdef #1{\the\secno}\writedef{#1\leftbracket#1}\wrlabeL{#1=#1}}
\def\subseclab#1{\xdef #1{\secsym\the\subsecno}%
\writedef{#1\leftbracket#1}\wrlabeL{#1=#1}}
\newwrite\tfile \def\writetoca#1{}
\def\leaderfill{\leaders\hbox to 1em{\hss.\hss}\hfill}
\def\writetoc{\immediate\openout\tfile=toc.tmp
   \def\writetoca##1{{\edef\next{\write\tfile{\noindent ##1
   \string\leaderfill {\noexpand\number\pageno} \par}}\next}}}

%
\catcode`\@=12 
%
\edef\tfontsize{\ifx\answ\bigans scaled\magstep3\else scaled\magstep4\fi}
\font\titlerm=cmr10 \tfontsize \font\titlerms=cmr7 \tfontsize
\font\titlermss=cmr5 \tfontsize \font\titlei=cmmi10 \tfontsize
\font\titleis=cmmi7 \tfontsize \font\titleiss=cmmi5 \tfontsize
\font\titlesy=cmsy10 \tfontsize \font\titlesys=cmsy7 \tfontsize
\font\titlesyss=cmsy5 \tfontsize \font\titleit=cmti10 \tfontsize
\skewchar\titlei='177 \skewchar\titleis='177 \skewchar\titleiss='177
\skewchar\titlesy='60 \skewchar\titlesys='60 \skewchar\titlesyss='60
\def\titlefont{\def\rm{\fam0\titlerm}
\textfont0=\titlerm \scriptfont0=\titlerms \scriptscriptfont0=\titlermss
\textfont1=\titlei \scriptfont1=\titleis \scriptscriptfont1=\titleiss
\textfont2=\titlesy \scriptfont2=\titlesys \scriptscriptfont2=\titlesyss
\textfont\itfam=\titleit \def\it{\fam\itfam\titleit}\rm}
 \ifx\answ\bigans\else scaled\magstep1\fi
\ifx\answ\bigans\def\abstractfont{\tenpoint}\else
\font\abssl=cmsl10 scaled \magstep1
\font\absrm=cmr10 scaled\magstep1 \font\absrms=cmr7 scaled\magstep1
\font\absrmss=cmr5 scaled\magstep1 \font\absi=cmmi10 scaled\magstep1
\font\absis=cmmi7 scaled\magstep1 \font\absiss=cmmi5 scaled\magstep1
\font\abssy=cmsy10 scaled\magstep1 \font\abssys=cmsy7 scaled\magstep1
\font\abssyss=cmsy5 scaled\magstep1 \font\absbf=cmbx10 scaled\magstep1
\skewchar\absi='177 \skewchar\absis='177 \skewchar\absiss='177
\skewchar\abssy='60 \skewchar\abssys='60 \skewchar\abssyss='60
\def\abstractfont{\def\rm{\fam0\absrm}
\textfont0=\absrm \scriptfont0=\absrms \scriptscriptfont0=\absrmss
\textfont1=\absi \scriptfont1=\absis \scriptscriptfont1=\absiss
\textfont2=\abssy \scriptfont2=\abssys \scriptscriptfont2=\abssyss
\textfont\itfam=\bigit \def\it{\fam\itfam\bigit}\def\footnotefont{\tenpoint}%
\textfont\slfam=\abssl \def\sl{\fam\slfam\abssl}%
\textfont\bffam=\absbf \def\bf{\fam\bffam\absbf}\rm}\fi
\def\tenpoint{\def\rm{\fam0\tenrm}
\textfont0=\tenrm \scriptfont0=\sevenrm \scriptscriptfont0=\fiverm
\textfont1=\teni  \scriptfont1=\seveni  \scriptscriptfont1=\fivei
\textfont2=\tensy \scriptfont2=\sevensy \scriptscriptfont2=\fivesy
\textfont\itfam=\tenit \def\it{\fam\itfam\tenit}\def\footnotefont{\ninepoint}%
\textfont\bffam=\tenbf \def\bf{\fam\bffam\tenbf}\def\sl{\fam\slfam\tensl}\rm}
\font\ninerm=cmr9 \font\sixrm=cmr6 \font\ninei=cmmi9 \font\sixi=cmmi6
\font\ninesy=cmsy9 \font\sixsy=cmsy6 \font\ninebf=cmbx9
\font\nineit=cmti9 \font\ninesl=cmsl9 \skewchar\ninei='177
\skewchar\sixi='177 \skewchar\ninesy='60 \skewchar\sixsy='60
\def\ninepoint{\def\rm{\fam0\ninerm}
\textfont0=\ninerm \scriptfont0=\sixrm \scriptscriptfont0=\fiverm
\textfont1=\ninei \scriptfont1=\sixi \scriptscriptfont1=\fivei
\textfont2=\ninesy \scriptfont2=\sixsy \scriptscriptfont2=\fivesy
\textfont\itfam=\ninei \def\it{\fam\itfam\nineit}\def\sl{\fam\slfam\ninesl}%
\textfont\bffam=\ninebf \def\bf{\fam\bffam\ninebf}\rm}
%
%

\hyphenation{anom-aly anom-alies coun-ter-term coun-ter-terms}
\def\inv{^{\raise.15ex\hbox{${\scriptscriptstyle -}$}\kern-.05em 1}}

\def\Dsl{\,\raise.15ex\hbox{/}\mkern-13.5mu D} 
\def\dsl{\raise.15ex\hbox{/}\kern-.57em\partial}

\font\bigit=cmti10 scaled \magstep1
\def\lspace{\ifx\answ\bigans{}\else\qquad\fi}
\def\lbspace{\ifx\answ\bigans{}\else\hskip-.2in\fi} 
\def\boxeqn#1{\vcenter{\vbox{\hrule\hbox{\vrule\kern3pt\vbox{\kern3pt
	\hbox{${\displaystyle #1}$}\kern3pt}\kern3pt\vrule}\hrule}}}
\def\mbox#1#2{\vcenter{\hrule \hbox{\vrule height#2in
		\kern#1in \vrule} \hrule}}  
%

\def\darr#1{\raise1.5ex\hbox{$\leftrightarrow$}\mkern-16.5mu #1}

\def\roughly#1{\raise.3ex\hbox{$#1$\kern-.75em\lower1ex\hbox{$\sim$}}}
\Title{HUTP-93/A006}{The hamiltonian reduction of the BRST complex and N=2
SUSY}
\centerline{Vladimir Sadov}
\bigskip\centerline{Lyman Laboratory of Physics}
\centerline{Harvard University}\centerline{Cambridge, MA 02138}
\centerline{and}\centerline{L.D. Landau Institute for Theoretical Physics,
Moscow}

\vskip .3in

We study the nonunitary representations of $N=2$ Super Virasoro algebra for the
rational central charges $\hat{c}<1$. The resolutions for the irreducible
representations of $N=2\ SVir$ in terms of the "2-d gravity modules" are
obtained and their characters are computed. The correspondence between $N=2$
nonunitary "minimal" models and the Virasoro minimal models + 2-d gravity is
shown at the level of states. We also define the hamiltonian reduction of the
BRST complex of $\widehat{sl}(N)/\widehat{sl}(N)$ coset to the BRST complex of
the W-gravity coupled to the W matter. The case $\widehat{sl}(2)$ is considered
explicitly. It leads to the presentation of $N=2$ Super Virasoro algebra as the
Lie algebra cohomology. Finally, we reveal the mechanism of the correspondence
$\widehat{sl}(2)/\widehat{sl}(2)$ coset --- 2-d gravity.

\Date{2/93}

\newsec{Introduction}

(The reader who is only interested in $N=2$ nonuninitary representations and
don't care for cosets for the first reading may skip the first four paragraphs
of the Introduction and the whole Section 2.)

In the first part of this paper we continue studying the relationships
between the
noncritical $W_N$ strings and
$\widehat{sl}(N)\widehat{sl}(N)_k/\widehat{sl}(N)_k$-cosets started in
\ref\SadI{V.~Sadov {\it On the spectra of
$\widehat{sl}(N)_k/\widehat{sl}(N)_k$-cosets and $W_N$ gravities.} Harvard
preprint HUTP-92/A055}.
Our main goal here is to explain the role of the hamiltonian reduction in the
story. It is well known, that the hamiltonian reduction \ref\BeO{M.~Bershadsky,
H.~Ooguri {\it Comm.~Math.~Phys. {\bf 126}(1992) 49}}, \ref\rFeFrI{B.~Feigin,
E.~Frenkel
{\it Phys.~Letts.~}{\bf B246}(1990) 75}  maps the
representations of
$\widehat{sl}(N)_k$ to that of $W_N$. (In a sense, we can consider this {\it as
a definition} of $W_N$). Technically speaking it is achieved by taking the
(semi-infinite) cohomology $M_W$ of $\hat{N_+}(sl(N))$ ("twisted" by some
character) of the given module $M_{sl(N)}$. So defined,  $M_W$ has a natural
structure of a
$W$- algebra module \rFeFrI.
When computing the spectrum of physical states in coset, we take the
(semi-infinite) cohomology of the {\it whole} $\widehat{sl}(N)$ acting on the
tensor product of two $\widehat{sl}(N)$ representations. (Usually one of them
is irreducible (the matter sector) and another is a Wakimoto representation
(Toda sector)).
There is a suspicion, backed by the explicit calculation for $\widehat{sl}(2)$
coset and $Vir$-gravity respectively, that the spectra of $\widehat{sl}(N)$
cosets and $W_N$ gravities must coincide \ref\AGS{ O.~Aharony, O.~Ganor,
N.~Sochen, J.~Sonnenschein, S.~Yankielowicz
{\it Physical states in G/G Models and 2d Gravity} TAUP-1961-92}
\ref\AGSI{O.~Aharony, J.~Sonnenschein, S.~Yankielowicz
{\it G/G Models and $W_N$ Strings} TAUP-1977-92}
\ref\Huyu{ H.~L.~Hu, M.~Yu
{\it  On BRST cohomology of SL(2)/SL(2) gauged WZNW models} AS-ITP-92-32}\SadI.
 To make the statement more precise, let us consider two
modules, $M_1$
and $M_2$ of  $\widehat{sl}(N)$ ($\widehat{sl}(2)$ in this example), and two
modules
$M_1^{DS}$ and $M_2^{DS}$ of $W_N$ corresponding
to the first pair by the (quantum Drinfeld-Sokolov) hamiltonian reduction.
For definiteness, let $M_1$ be an "admissible" irreducible representation
($k_1+N={p\over q}$ ---
rational )
and $M_2$ be a Wakimoto ( free fields ) representation with the value of the
central charge $k_2=-k_1-2N$.

Then, the $\widehat{sl}(2)$-BRST homology of $M_1 \otimes M_2$ do coincide with
$W_2=Vir$-BRST homology of  $M_1^{DS} \otimes M_2^{DS}$. The similar result is
true if we take two Wakimoto or one Wakimoto and one "transposed" Wakimoto
modules for $\widehat{sl}(2)$ and two free boson Fock modules for $Vir$
respectively.

Whereas the spectra of cosets can be found explicitly, it is quite
difficult to do for
 $W_N$-gravity. Although it is really possible to construct
a
BRST complex in that case \ref\BNW{W.~Lerche, D.~Nemeshansky,M.~Bershadsky,
N.~Warner
{\it A BRST Operator for non-critical W-Strings}  HUTP-A034/92}
\ref\BSS{E.~Bershgoeff, A.~Sevrin, X.~Shen
{\it A derivation of the BRST operator for noncritical strings} preprint}, it
is not at all clear why it is possible.
Then, it
appears that this complex is not very convenient for the direct computations.

 Taken together, all these facts motivate a desire to define a procedure of
hamiltonian reduction not only for one $\widehat{sl}(N)$ module (quantum
Drinfeld-Sokolov reduction), but also for the
whole BRST complex.
 In Section 2 we address this
issue and give a proper modification of the reduction procedure.
Then we outline how it works for $\widehat{sl}(2)$.

At this stage, quite naturally, the (topologically twisted) $N=2$
superconformal algebra appears. ($N=2\ SVir$ for the basic example of
$\widehat{sl}(2)$.) We show that the "reduced" $\widehat{sl}_k(2)$-BRST
complex, with an irreducible representation in the matter sector is just a
direct sum of two copies of the irreducible representation of $N=2\ SVir$ with
the central charge $\hat{c}={k\over k+2}$ as a vector space. The differential
is given by the zero mode $G_0^+$ of the superconformal current. (This is
proven in Section 4.1).

In Section 3 we present the necessary background material on the nonunitary
representations of $N=2\ SVir$, following \ref\SadII{V.~Sadov {\it Free field
resolution for nonunitary representations of N=2 SuperVirasoro} Harvard
preprint HUTP-92/A070}. In fact, in \SadII  was considered only the case of
irrational $\hat{c}$, so here we have to generalize that results to the more
complicated situation of rational $\hat{c}$. We obtain a resolution for the
irreducible representations of $N=2\ SVir$ in terms of the products
$L(Vir)\otimes F(Liouv.)\otimes F_{gh}$ (a "2-d gravity" resolution), where
$L(Vir)$, $F(Liouv.)$ and $F_{gh}$ denote respectively a Virasoro irreducible
representation ("the matter"), a free bosonic Fock space ("the Liouville
field") and a two fermions Fock space ("the diffeomorphisms ghosts"). Using
this resolution we find in particular a character formula for the irreducible
representations of $N=2\ SVir$.

 This formula explicitly incorporates the Lian-Zuckerman states of the Virasoro
$(p,q)$  (with ${p\over q}=k+2$) minimal model coupled to gravity. Thus it
explains {\it for the representations} the relations between $N=2\ SVir$ and
2d-gravity, found in \ref\SeBGR{B.~Gato-Rivera, A.~Semikhatov {\it
Phys.~Letts.}{\bf B293} (1992) 72},\ref\BLNW{M.~Bershadsky, W.~Lerche,
D.~Nemeshansky, N.~Warner {\it N=2 Extended superconformal structure of Gravity
and W Gravity coupled to Matter}  HUTP-A034/92} {\it for the chiral algebras}.
It also gives a piece of evidence in favour of the idea \ref\Los{A.~Losev {\it
Descendants constructed from matter field and K.~Saito higher residue pairing
in Landau-Ginzburg theories coupled to topological gravity} preprint
TPI-MINN-92-40-T} that $N=2$ minimal theory "may already know about 2-d
gravity". Also it is nice to have an object which puts together an infinite
number of LZ states for the given matter field.

Finally, in Section 4.2 we combine the BRST complex hamiltonian reduction of
Sec. 2 and the resolutions of Sec. 3 to trace explicitly the mechanism
identifying the physical states in $\widehat{sl}(2)$ coset and 2-d gravity.
{}From the point of view of the $N=2\ SVir$ representations theory this last
section may be viewed as the consistency check for the results we obtained in
the Section 3, the "2-d gravity resolution" in particular.

\newsec{Hamiltonian reduction of $\widehat{sl}(N)$ BRST complex}
\subsec{Some motivations}
 In a sense, this section is the second Introduction. We wish to informally
explain here what we mean by the hamiltonian reduction of the BRST complex. The
technical details can be found in the next two sections.
First, it seems natural to repeat some motivations from \SadI.

In \BeO,\rFeFrI the quantum Drinfeld-Sokolov (DS) hamiltonian reduction for
Wakimoto and irreducible representations
has been defined as the homology $H^0$ of the DS BRST complex, associated with
the constraints
\eqn\erecon{\eqalign{
&e^\alpha (z)=1\ if\ \alpha\ is\ a\ simple\ root \cr
&e^\alpha (z)=0\ if\ \alpha \ is\ not\ a\ simple\ root \cr
}}
Suppose now that we have a $\widehat{sl}(N)$-BRST complex which computes the
cohomology of the tensor product of two modules $M_1$, $M_2$. Here $M_1$ can be
either an irreducible representation $L_k(\widehat{sl}(N))$ or a Wakimoto
representation $Wak_k$ (or possibly a "transposed" Wakimoto representation
$\widehat{Wak}_k$) and $M_2$ is a Wakimoto representation. In principle, it is
also interesting to consider other combinations. In the language of the
$\widehat{sl}(N)/\widehat{sl}(N)$ coset model the modules $M_1$, $M_2$
represent respectively the fields of{\it the matter} and of {\it the
Liouville-Toda} sectors of the theory.
The cohomology $H^*_{Q_{BRST}}(M_1\otimes M_2)$, where $M_1$ runs over some
specified set of representations with the fixed level $k$ and $M_2$ runs over
all Wakimoto representations with level $-2N-k$ form {the spectrum of physical
states} of the theory. Usually we consider rational levels $k+N={p\over q}$ and
  restrict $M_1$ to the "admissible" irreducible representations of
$\widehat{sl}(N)$.

We wish to have for {\it the whole BRST complex} something like what the DS
reduction is for the single module. More precisely, ultimately we wish to
obtain a $W_N$ BRST complex by this "something".
{}From the first sight it seems natural to try reduction independently on each
factor. So we would have to add two sets of the DS ghost-antighost pairs,
labeled by the positive roots of ${sl}(N)$ and consider the DS BRST operators
$Q_1$, $Q_2$ acting respectively on $M_1$ and $M_2$, then take the product
$M_1^{DS}\otimes M_2^{DS}\otimes \{\widehat{sl}(N)\ ghosts\}$ where
$M_i^{DS}=H^0_{Q_i}(M_i\otimes \{DS\ ghosts\}_i)$ are the reduced modules.
$M_i^{DS}$ are the representations of $W_N$ algebra.

This procedure does not work because we have to require that the
reduction BRST operator, which is $Q_1+Q_2$ here, commute with
$\widehat{sl}(N)$-BRST operator, and it is not difficult to
check,
that there is no proper modification of $Q_1+Q_2$, commuting with $Q_{BRST}$
\foot{mainly because in $Q_1+Q_2$ necessarily participate not only
the symmetric combinations of currents like $J_1^a+J_2^a$ but also the
antysymmetric ones like $J_1^a-J_2^a$.
The commutators of the latter with $Q_{BRST}$ cannot
be compensated by adding extra terms with ghosts}.
The other problem to deal with is what to do with the $\widehat{sl}(N)$ ghosts.
They form a representation of $\widehat{sl}(N)$ (at level $2N$) and the general
ideology requires to reduce this representation also. (We must somehow obtain
the $W_N$ ghosts!) After a short reasoning it
seems very natural {\it not} to introduce special reduction ghosts at all and
to try to
make the reduction of the BRST complex using the $\widehat{sl}(2)$ ghosts
themselves.
\subsec{The basic definition}
(If the reader is not interested in "general nonsense" and only wants to
consider $\widehat{sl}(2)$ and 2-d gravity, he may skip this subsection and
start from 2.3.)

Let us give a definition.

{\bf Definition-Hypothesis.} {\it Let $M_1$ be either the irreducible or the
Wakimoto representation at level $k$, and $M_2$ be the Wakimoto representation
at level $-k-2N$ of the $\widehat{sl}(N)$ algebra.
Then there exists a spectral sequence, converging (at the second term) to the
$\widehat{sl}(N)$-BRST cohomology $H^*_{Q_{BRST}}(M_1\otimes M_2)$ of the
module $M_1\otimes M_2$. Denote by ${Q_R}$ a differential in the  first term of
the spectral sequence. Then the second term of the spectral sequence, i.e. the
complex  ($H_{Q_R}^*$, $Q_W$) with the
differential $Q_W$ is quasiisomorphic (has the same cohomology ) to the
$W_N$-BRST complex of the module  $M_1^{DS}\otimes M_2^{DS}$ where the
superscript "DS" denotes the standard (Drinfeld-Sokolov) hamiltonian reduction.

We say, that $Q_R$ "makes the hamiltonian reduction of $\widehat{sl}(N)$-BRST
complex" and call $(H_{Q_R}^*,Q_W)$ "the reduced BRST complex".}

{\bf Comment} Because of the required quasiisomorphism the cohomology computed
by the spectral sequence are determined by the cohomology
of $W_N$-BRST. One the other hand this is the $\widehat{sl}(n)$-BRST cohomology
just by the definition. Thus we see that the equivalence of spectra of cosets
and W-gravity should follow from the hypothesis above.

\subsec{An example of the construction.}
We have formulated a general hypothesis. It may seem a little bit complicated
and not very explicit.
Now we wish to show how it can be proved for the algebra $\widehat{sl}(2)$. In
this example the coincidence of spectra was known for some time from the
explicit computation, which was fairly straightforward in this case. However we
shall see that our construction  is nontrivial already in this simplest case
and gives rise to the interesting N=2 supersymmetric structure.

Let us we consider a decomposition $Q_{BRST}=Q_R+Q_W$ with
\eqn\esuII{\eqalign{
&Q_R=\oint c^+(E_1+E_2)+c^0(H_1+H_2+2(b_+c^+ - b_-c^-)) \cr
&Q_W=\oint c^-(F_1+F_2+c^+b_-) \cr
}}

It is convenient to rephrase our hypothesis in this particular case in the form
of the

{\bf Theorem.} {\it Let $M_1$ be either the irreducible or the  Wakimoto
representation at level $k$, and $M_2$ be the Wakimoto representation at level
$-k-4$ of the $\widehat{sl}(2)$ algebra.
Then $\widehat{sl}(2)$-BRST complex for $M_1\otimes M_2$ has a structure of the
double complex with differentials  $(Q_R,Q_W)$ such that  $Q_R+Q_W=Q_{BRST}$.
Consider a spectral sequence of this double complex whose first term has a
differential ${Q_R}$. Then the second term of the spectral sequence, i.e. the
complex  ($H_{Q_R}^*$, $Q_W$) with the
differential $Q_W$ is quasiisomorphic (has the same cohomology ) to the
Virasoro-BRST complex of the module  $M_1^{DS}\otimes M_2^{DS}$ where the
superscript "DS" denotes the standard (Drinfeld-Sokolov) hamiltonian reduction.

In other words, there exists a hamiltonian reduction of  $\widehat{sl}(2)$-BRST
complex".}

It is a straightforward calculation to check that the operators \esuII do
define the structure of the double complex on the $\widehat{sl_k}(2)$ BRST
complex. What is $H_{Q_R}^*(\widehat{sl_k}(2)\otimes Wakim_{-4-k})$ --- the
$Q_R$ cohomology of the chiral algebra  $\widehat{sl_k}(2)\otimes Wakim_{-4-k}$
here?

Doing the direct computation\foot{
It can be a good idea to use the Matematica OPE package \ref\rOPE{K.~Thielemans
{\it Int. J. Mod. Phys.} {\bf C} Vol.2 No.3, 787 (1991)} to do this!}
one convinces oneself that two currents
\eqn\ecohI{\eqalign{
&G^+(z)={1 \over 2(k+2)}\bigl(c^-(F_1+F_2+c^+b_0)+2\partial (\gamma c^-)\bigr)
\cr
&G^-(z)=b_-(E_1-E_2) \cr
}}
belong to $H_{Q_R}^*$. Their OPE is
\eqn\eOPEpm{
G^+(z)G^-(0) = {{k/k+2} \over z^3}+ {J(z) \over z^2}+{{T(z)+\partial J(z)}
\over z}+regular\ terms
}
where
\eqn\eJ{\eqalign{
&J(z)=:c_-b^-(z)-{2 \over \sqrt{2(k+2)}}\bigl( \partial \varphi (z)+{1\over
\sqrt{2(k+2)}}([E_1+\beta ]\gamma +:b_+c^+:)\bigr) \cr
&+{1\over {2(k+2)}}\{Q_R,b_0 \} \cr
}}
The operators $J(z)$, $T(z)$   also are the nontrivial elements of
$H_{Q_R}^*(\widehat{sl_k}(2)\otimes Wakim_{-4-k})$. Moreover, it is true that
$T(z)$ is equivalent modulo $Q_R$-exact term to the twisted stress-energy of
the coset and that four currents $G^+(z),\ G^-(z),\ T(z), \ J(z)$    form a
closed chiral algebra which is just a (topologically twisted) N=2 SuperVirasoro
with the central charge
\eqn\ecech{ \hat{c}={k \over k+2}}
This is not very surprising. For example, the similar phenomena was observed in
  \ref\rMuVa{S.~Mukhi, C.~Vafa {\it Two dimensional Black hole, c=1
Non-Critical Strings and a Topological Coset Model. } Harvard preprint
HUTP-93/A002} for the Kazama-Suzuki coset with $k=-3$ in our notations.

 Thus $SVir \subset H_{Q_R}^*(\widehat{sl_k}(2)\otimes Wakim_{-4-k})$.
In fact, there is also an operator $c^0_0$ --- the zero mode of the ghost
$c^0(z)$ --- which belongs to $H^*_R$. In the next sections we shall see using
more complicated technique that actually
\eqn\ecca{
H_{Q_R}^*\bigl(\widehat{sl_k}(2)\otimes Wakim_{-4-k}\bigr)=\bigl[N=2\
SVir)\oplus c^0_0(N=2\ SVir\bigr]}
as a {\it chiral algebra}. For now let us assume this is true.

Similarly it can be shown
(see Sec.4.1) that {\it for the representations} the cohomology
$H_{Q_R}^*(L_k(\widehat{sl}(2))\otimes Wak_{-4-k})$ are given just by the
direct sum of two copies (one is again shifted by $c_0^0$) of the irreducible
representation $L(N=2\ SVir)$ of N=2 $SVir$:
\eqn\ecrr{
H_{Q_R}^*\bigl(L_k(\widehat{sl_k}(2))\otimes Wak_{-4-k}\bigr)=\bigl[L(N=2\
SVir)\oplus c^0_0L(N=2\ SVir)\bigr]}

The crucial for the following observation is that the second differential $Q_W$
of our double complex is just a zero mode of the superconformal currents:
\eqn\eDpr{Q_W=G^+_0}
Thus the the {\it reduced BRST complex} for $\widehat{sl_k}(2)$
can be expressed solely in terms of N=2 $SVir$:
\eqn\esured{\bigl(H_{Q_R}^*(L_k(\widehat{sl_k}(2)\otimes
Wak_{-4-k}),{Q_W}\bigr)=\bigl(L(N=2\ SVir)\oplus c^0_0L(N=2\ SVir),G_0^+\bigr)
}
Now we shall use the relation between N=2 $SVir$ and 2d gravity coupled to the
minimal matter found in \SeBGR, \BLNW  to establish the quasiisomorphism the
main hypothesis claims. Let me remind here this relation.

Consider the BRST complex of 2d gravity coupled to the minimal matter with the
central charge $c_M$. The chiral algebra of the matter sector is $Vir$, and
that of the Liouville and ghost sectors are correspondingly $Heis$ and $Clif$
with the total central charge $-c_M$. Then

{\it There is an embedding
\eqn\emap{N=2\ SVir \rightarrow Vir\otimes Heis \otimes Clif}
of the chiral algebras and the corresponding map on the representations such
that $G^-(z)\rightarrow b(z)$, $G^+(z)\rightarrow J_{BRST}(z)$, where
$J_{BRST}(z)$ is a Vir-BRST current plus a total derivative term. In
particular, $G^+_0 \rightarrow Q_{Vir}$ --- a BRST operator of 2-d gravity.}

We shall show in Sections 3,4 that the similar relation exists at the level of
representations. In particular, for the irreducible representation $L(N=2\
SVir)$ there exists a resolution in terms of the "2-d gravity" modules
$L(Vir)\otimes F(Liouv.)\otimes F_{gh}$, where $L(Vir)$, as usual, denotes the
irreducible representation (of Virasoro, this time), $F(Liouv.)$ is a free
boson Fock space and $F_{gh}$ is the diffeomorphisms ghosts Fock space:
\eqn\ereslI{\eqalign{
&0\rightarrow L(N=2\ SVir)\rightarrow L(Vir)\otimes F(Liouv.)\otimes F_{gh}
\rightarrow L'(Vir)\otimes F'(Liouv.)\otimes F_{gh}\rightarrow \cr
&\rightarrow L''(Vir)\otimes F''(Liouv.)\otimes F_{gh}\rightarrow \cdots
}}
(This resolution is infinite in the most interesting cases, see Section 3.)

To complete the proof, we should substitute this resolution into the reduced
BRST complex \esured to obtain the double complex of "2-d gravity" modules with
two differentials. One of them comes from the resolution \ereslI. The other one
is just
\eqn\ediff{
Q_W=G^+_0=Q_{Vir}
}
 We are almost done now. The cohomology of the double complex are computed in
the Sec. 4.2 where it is shown that they reduce to $H^*_{Q_{Vir}}(L(Vir)\otimes
F(Liouv.)\otimes F_{gh})$ --- the 2-d gravity BRST cohomology of the first term
in the resolution \ereslI.
Before passing to the details of the proof let us summarize what we have
learned and propose  the possible generalization for the algebras
$\widehat{sl}(N)$.
First, we decomposed the Lie algebra BRST operator into two pieces,
corresponding to the Borel subalgebra of $sl(2)$ and its compliment (+some
ghost terms corrections) to obtain a structure of a double complex. This
decomposition for $\widehat{sl}(N)$ goes through if we
take as a $Q_R$ a proper piece of $\widehat{sl}(N)-Q_{BRST}$, corresponding to
the
maximal parabolic subalgebra of (the finite dimensional) ${sl}(N)$.

Then we computed the cohomology of $Q_R$ which turned to be the irreducible
representation of N=2 $SVir$. For $\widehat{sl}(N)$ it is likely the
representation of N=2 SuperW algebra. The cohomology has the structure of a
complex with differential
$Q_W=G^+_0$.

Finally we used the map \emap \SeBGR, \BLNW to relate this complex to the BRST
complex of 2d gravity coupled to the minimal matter. Such map into W-gravity+W
minimal matter also exists for any N=2 SuperW algebra \BLNW. Then we show that
the cohomology of the reduced BRST complex are the same as of the W-system.

\newsec{Representations of N=2 SuperVirasoro}
\subsec{The bosonisation formulas.}
 In this subsection we recall the results from the representation theory of
$N=2 SVir$, obtained in \SadII. It is important, that we actually need the {\it
nonunitary } representations of $N=2 SVir$, it follows from the formula \ecech
for the central charge (remember that $k$ is just rational, not necessarily
integer).

Let us introduce the basic notations.
We have a system consisting of the Virasoro algebra ($Vir$)
--- a matter sector, a free bosonic field $\phi$ with the background charge
($Heis$) --- a Liouville sector and a pair of fermions $b,c$ of spins 2,-1
($Clif$) --- the diffeomorphism ghosts ( in the brackets are the names of the
corresponding chiral algebras). We require the total central charge be equal to
zero. Then the currents
\eqn\eNtwo{\eqalign{&J(z)=:cb:+\alpha _-\partial \phi \cr
&G^+(z)=:c[T_{Vir}+T_\phi +{1 \over 2}T_{bc}]:-2\alpha _-\partial (c\partial
\phi ) +{1 \over 2}(1-2\alpha _-^2)\partial ^2c \cr
&G^-(z)=b(z) \cr
&T=T_{Vir}+T_\phi +T_{bc} \cr
&T_\phi =-{1 \over 4}:(\partial \phi )^2:+\beta _0 \partial ^2 \phi \cr
}}

satisfy the OPE of (twisted) N=2 $SVir$  chiral algebra. In fact these formulas
give the embedding of N=2 $SVir$  into the tensor product of three other chiral
algebras
$Vir \otimes Heis \otimes Clif$ ( we used this fact in the previous section ).
To describe the properties of this map  it is technically convenient to
bosonise the Virasoro algebra by the free field $X$ with the background charge
$\alpha _0$. In other words we embed $Vir$ into the Heisenberg algebra
generated by $\partial X$ which we denote by $Heis'$ to distinguish it from the
Liouville $Heis$. Substituting the bosonized matter stress energy
\eqn\ebos{\eqalign{&T_{Vir}(z) ={1 \over 4}:(\partial X )^2:+\alpha _0 X \cr
&\beta _0^2-\alpha _0^2=1, \ \ \alpha _{\pm}=\alpha _0 \pm \beta _0 \cr
&c_{Vir}=1-24\alpha _0^2 \cr
}}

into \eNtwo
we finally obtain the bosonisation prescriptions for N=2 $SVir$ we need.
Unlike the standard bosonisation \ref\rUniMSS{ G.~Mussardo, G.~Sotkov,
M.~Stanichkov {\it Int.~J.~Mod.~Phys}{\bf A4} (1989) 1135},\ref\rUniIto{
K.~Ito, {\it Nuclear Physics}{\bf B332} (1990) 566},
\ref\Chung{C.~Ahn,S.~Chung, S.~H.~H.~Tye ``New Parafermion, SU(2) Coset and N=2
Superconformal Field Theories'',
 {\it Nucl. Phys.} {\bf B365}, 191-242 (1991) }the formulas for $G^+(x)$ and
$G^-(z)$ are very asymmetric.
Comparing \ebos with \ecech we see that
\eqn\ealmi{\alpha _-={-1 \over \sqrt{k+2}}
}
( it is a nonstandard notation for $\widehat{sl}(2)_k$!).
For the representations, we take a Fock space
\eqn\defo{
F_{\alpha \beta}=F_{\alpha}(X) \otimes F_{\beta}(\phi ) \otimes F_{gh}
}
$F_{\alpha}(X)$ and $F_{\beta}(\phi )$ here are the standard Fock modules of
$Heis'$ and $Heis$  with vacuums $|\alpha >$ and $|\beta >$  respectively and
$F_{gh}$ is a ghosts Fock space (a $Clif$ Verma module) with the vacuum vector
$|0>$ annihilated by
\eqn\evac{c_n|0>=0 \ n > 1, \ \ b_n|0>=0 \ n>-2 }

In $F_{gh}$ we take a vector $|0>_{phys}=c_1|0>$ and define the N=2 vacuum as
$\Omega =|\alpha > \otimes |\beta > \otimes |0>_{phys}$. This procedure is well
known in string theory. Here we use it to endow the free field Fock space
$F_{\alpha \beta}$ with a structure of a highest weight N=2 $SVir$  module:
\eqn\eNtwovac{\eqalign{&L_n\Omega =J_n\Omega =G_n^-\Omega =0, \ n\geq 0 \cr
&G_n^+\Omega =0, \ n>0 \cr
&L_0\Omega =\Delta \Omega =(-1+\alpha (\alpha -2\alpha _0)-\beta (\beta -2\beta
_0)) \cr
&J_0\Omega =Q \Omega=(1+2\alpha _-\beta )\Omega \cr
}}

It is convenient to rewrite the formula for the conformal weight as
\eqn\ewei{\Delta (\alpha ,\beta )=(\alpha +\beta -\alpha _+)(\alpha -\beta
-\alpha _-)
}
There are two {\it screening operators} in our bosonisation. One of them is
just $E=\oint :e^{\alpha _+X(z)}:$. It comes from the bosonisation of the $Vir$
matter. The other one is $F_1=\oint :b(z)e^{-{\alpha _+ \over 2}(X(z)+\phi
(z))}:$  \BLNW,\ref\rRoz{L.~Rozansky  {\it a letter to M.~Bershadsky},
1989},\ref\rDots{Vl.~Dotsenko {\it Mod.~Phys.~Lett.}{\bf A7} (1992) 2505}. It
is fermionic and local to itself:
\eqn\efnilp{F_1^2=0 }

Together, $E$ and $F_1$ form a quantum superalgebra $u_q(n_+(sl(2|1)))$ with
$q=e^{\pi i \alpha _+^2}$. Namely they satisfy the Serre relation:
\eqn\eSerr{E^2F_1-(q+q^{-1})EF_1E+F_1E^2=0}

(As usual, the left hand side of \eSerr is to be understood as a part of some
formal polynomial in screenings acting on the appropriate state.)

There is a simple but important remark to me made here. The basic object for
the correspondence N=2 $SVir \rightarrow \{2-d\ gravity\}$ above is the
Virasoro algebra itself, not the free field $X(z)$ which is just a useful tool
for describing this correspondence.
Bosonising the Virasoro algebra, we could choose a screening operator
$E^{(-)}=\oint :e^{\alpha _-X(z)}:$ instead of $E^{(+)}=\oint :e^{\alpha
_+X(z)}:$. The vertex operators corresponding to $E^{(-)}$, $F$ are local with
respect to each other, so we may simply set
\eqn\ecomm{ E^{(-)}F-FE^{(-)}=0
}
Hence the algebraic structure produced by the pair $E^{(-)}$, $F$ is much
simpler and therefore much less powerful for the purposes of the representation
theory then the structure produced by $E^{(+)}$, $F$. Therefore in this
subsection we'd better stick to the latter. In fact, we will be able to take
advantage of the simplicity of \ecomm later, {\it when we already know the
representation theory of} N=2 $SVir$.

Using very rigid conditions \efnilp, \eSerr we can classify the irreducible
representations of N=2 $SVir$ according to the types of the free fields
resolution they have. By such resolution we mean the complex of the free field
Fock spaces with the cohomology being nontrivial only in the zero degree where
it is represented by the irreducible representation. Having a complex we
compute its Euler characteristics (character valued, as usual) which turns to
be a character of the irreducible $L_{\alpha \beta }$. It is convenient to deal
with the normalized characters
\eqn\eintrep{\eqalign{
&\tilde{\chi}_{\alpha \beta}={\chi _{\alpha \beta} \over \chi (F_{\alpha ,
\beta})}\ , \ \chi (F_{\alpha , \beta})=Tr_{F_{\alpha ,
\beta}}(q^{L_0}x^{2J_0}) \cr
}}
\subsec{The case of irrational $\hat{c}$}
First let me describe the classification for the generic (irrational) values of
$\hat{c}$. Depending on $(\alpha ,\beta )$, the irreducible representation
$L_{\alpha ,\beta }$ may belong to either of the four following types.

{\bf Case I}. $(\alpha , \beta )$ is generic, the module $F_{\alpha \beta }$ is
irreducible. The corresponding complex is therefore trivial,
$\tilde{\chi}_{\alpha \beta}=1$.

{\bf Case II.}
\eqn\econdII{\alpha _{n\ m}= \alpha _+{1-n \over 2}+\alpha _-{1-m \over 2} }
$\beta$ is generic. This case essentially reduces to the well known theory for
the Virasoro algebra. The map $E^n$ is surjective, its kernel is a submodule in
${F_{\alpha  \beta}}$ generated by the highest weight vector. It gives
\eqn\rescsII{
L(N=2\ SVir)=L_{m\ n}(Vir)\otimes F_{\beta}\otimes F_{gh}
}
 This is the only map {\it from} $F_{\alpha  \beta}$ and the only map {\it to}
$F_{\alpha +n\alpha _+\ \beta}$; there is no maps {\it to} $F_{\alpha  \beta}$
or {\it from} $F_{\alpha _+\alpha _+\ \beta}$. Therefore the submodule in
$F_{\alpha  \beta}$ generated by the highest weight vector is irreducible. The
quotient of $F_{\alpha  \beta}$ by this submodule is also irreducible and
coincides with $F_{\alpha +n\alpha _+\ \beta}$. The character is
\eqn\echarII{\tilde{\chi}_{\alpha \beta}=1-q^{nm}}
This was the typical example of the argument to be used in this sort of
constructions.

{\bf Case $III_-$.}
\eqn\econdIIIminus{\alpha -\beta =-\alpha _-l }
When $l \geq -1$ the map $F_1$ sends the highest weight vector of ${F_{\alpha ,
\beta}}$ to a nonzero element, which generate in $F_{\alpha -{\alpha _+\over
2}, \beta -{\alpha _+\over 2}}$ a proper submodule $SF_{\alpha -{\alpha _+\over
2}, \beta -{\alpha _+\over 2}}$. There is no other maps into $F_{\alpha
-{\alpha _+\over 2}, \beta -{\alpha _+\over 2}}$, so $SF_{\alpha -{\alpha
_+\over 2}, \beta -{\alpha _+\over 2}}$  is the only proper submodule.
Therefore it must coincide with the kernel of the map $F_1$ from $F_{\alpha
-{\alpha _+\over 2}, \beta -{\alpha _+\over 2}}$ to $F_{\alpha -{\alpha _+},
\beta -{\alpha _+}}$. This means that the diagram $III_-$ is {\it exact} ---
the image of the incoming arrow coincide with the kernel of the outgoing arrow.

The diagram $III_-$ has already a natural structure of the complex. The graded
components are just the Fock spaces at the vertices and the differentials are
given by the arrows. This complex is infinite in both directions. It is exact,
so its cohomology is trivial. To obtain the resolution of $L_{\alpha \beta }$
one cuts the diagram by the arrow going {\it from} $F_{\alpha \beta }$ to
obtain two complexes with the equal cohomology (so there are two resolutions in
fact), one can use either of them. The character is
\eqn\echarIIIminus{\eqalign{
&\tilde{\chi}_{\alpha \beta}={1 \over 1+x^{-1}q^{l+1}} \cr
}}
We see that for $l \geq -1$ the formula \echarIIIminus can naturally be
interpreted as a character of the representation with the highest weight
$(\Delta _{\alpha \beta},q_{\alpha \beta})$. But for $l <-1$ the identical
transformation
\echarIIIminus $\rightarrow {xq{-l-1} \over 1+xq^{-l-1}}$ shows that the
character we compute now correspond to the weight  $(\Delta _{\alpha +{\alpha
_+ \over 2}, \beta +{\alpha _+ \over 2}},q_{\alpha +{\alpha _+ \over 2},\beta
+{\alpha _+ \over 2}})$. The reason for this phenomena is simple. For $l <-1$
the map $F_1$ kills the highest weight vector of $F_{\alpha \beta }$ and sends
some vector $w \in F_{\alpha \beta }$ to the highest vector of $F_{\alpha
-{\alpha _+\over 2}, \beta -{\alpha _+\over 2}}$. Hence each Fock space has one
cosingular vector $w_{\alpha \beta}$ and the irreducible representation is a
{\it submodule} of $F_{\alpha \beta }$ generated by the highest weight vector
of the Fock space\foot{Compare to $l \geq -1$ when the irreducible
representation was a {\it quotient } of the Fock space.}. Now, to obtain a
resolution of $L_{\alpha \beta }$ we should cut the diagram $III_-$ by the
arrow coming {\it into} $F_{\alpha \beta }$. The character is
\eqn\echarIIIplus{\eqalign{
&\tilde{\chi}_{\alpha \beta}={1 \over 1+xq^{-l-1}} \cr
}}

{\bf Case $III_+$}
\eqn\econdIIIpluplu{\alpha +\beta=\alpha _-(l+1)+\alpha _+}

One can repeat everything that have been said about $III_-$. The only subtlety
here is to check that the composition of two consequent maps in the diagram is
zero. The reader should convince oneself it is true using \eSerr and simple
$q$-polynomial identities. The formulas for the characters are given by the
same formulas \echarIIIminus ,\echarIIIplus.

{\bf Case $IV_-$} --- the conditions II and $III_-$ ($IV_-$) are met
simultaneously.
\eqn\econdIV{\eqalign{
&\alpha _{nm}= \alpha _+{1-n \over 2}+\alpha _-{1-m \over 2} \cr
&\beta _{nml}= \alpha _+{1-n \over 2}+\alpha _-{1-m+2l \over 2} \cr
}}
 We denote $F_{\alpha _{nm}\beta _{nml}}$ by $F_{nml}$. We know everything
already about the maps in the diagram. First consider the resolution of the
representation with $n\geq 0$, i.e. belonging to the left column in Fig.1. To
obtain a resolution we should again cut the diagram by the horizontal line
crossing the arrow  {\it above} $(\alpha \beta )$ for $l\geq -1$  or {\it
below} $(\alpha \beta )$  for $k<-1$. Keeping the upper half we end up with a
"ladder" shown in the Fig.2.
The structure of the complex $\{C^r, d_{(r)}\}_{r\geq 0}$is given by
\eqn\ecompIV{\eqalign{
&C^0=F_{n+1\ m\ l},\ C^r=F_{n+1+r\ m\ l}\oplus F_{-(n+r)\ m\ l} \cr
&d_{(0)}=E^{n+1}\oplus F_1, \ d_{(r)}= \pmatrix{&F_1 &0 \cr
					      &E^{n+1+r} &x_{n+r}EF_1-F_1E \cr}
}}
\eqn\ecreq{x_{l+1}=(q+q^{-1})-{1\over x_l},\ \ x_0=q+q^{-1}}
 The character is
\eqn\echarIV{\eqalign{
&l\geq -1 \ \ \tilde{\chi}_{\alpha \beta}={1-q^{m(n+1)}+q^{m-l-1}(1-q^{mn})
\over (1+x^{-1}q^{l+1})(1+xq^{m-l-1})} \cr
&l<-1 \ \ \tilde{\chi}_{\alpha \beta}={1-q^{mn}+q^{m-l-1}(1-q^{m(n-1)}) \over
(1+xq^{-l-1})(1+xq^{m-l-1})} \cr
}}

{\bf Case $IV_+$} --- the conditions II and $III_+$ are met simultaneously.
\eqn\econdIV{\eqalign{
&\alpha _{nm}= \alpha _+{1-n \over 2}+\alpha _-{1-m \over 2} \cr
&\beta _{nml}= \alpha _+{1+n \over 2}+\alpha _-{-1+m+2(l+1) \over 2} \cr
}}
One just repeats what was said about $IV_+$ (probably it is better to take a
bottom half of the cut diagram to construct a resolution of the representations
with $n>0$ in this case).  The formulas for the characters \echarIV are
applicable.

Note that the equation
\eqn\ezerwei{\Delta (\alpha ,\beta )=(\alpha +\beta -\alpha _+)(\alpha -\beta
-\alpha _-)=0
}
has two branches of solutions corresponding to either $III_\pm$ or $IV_\pm$
with $l=-1$.

It is important to note that the resolutions in the cases $I-IV$ above can be
rewritten in terms of modules $L_\alpha \otimes F_\beta \otimes F_{gh}$ instead
of $F_\alpha \otimes F_\beta \otimes F_{gh}$, where $L_\alpha$ is the
irreducible representation of the Virasoro algebra.

Indeed, for the case II we have shown this explicitly in \rescsII. For the
cases $I,III$ this is trivial because the irreducible and free field
representations are the same object: $L_\alpha=F_\alpha$. In the case $IV$ we
can compute the cohomology of the free field resolution using the spectral
sequence, associated with the "vertical" filtration, shown in Fig.2. The first
term of this spectral sequence computes the "horizontal" (in Fig.2) cohomology,
which gives exactly $L_\alpha \otimes F_\beta \otimes F_{gh}$. Then the second
term represents the resolution we are after:
\eqn\aures{
0\rightarrow L_{nml}\rightarrow L_{nm}\otimes F(\beta _{nml})\otimes F_{gh}
\rightarrow L_{n+1\ m}\otimes F(\beta _{n+1\ ml})\otimes F_{gh}\rightarrow
\cdots
}
Existence of such resolution in terms of irreducible representations of
Virasoro is not surprising at all, because it is the Virasoro algebra, not a
free field $X(z)$ which is basic in the correspondence between Supervirasoro
and 2-d gravity, so everything has to be expressible in terms of it. This fact
should be viewed as a counterpart for the {\it representations} of the map
\emap between the {\it chiral algebras}.

In such form, the resolutions above can  be generalized to the rational values
of $\hat{c}$.

\subsec{The case of rational $\hat{c}$}
Up to now we dealt with irrational $k$. But as \ecech shows, to consider the
most interesting "minimal" cosets we must take $k$ rational! The free field
resolution becomes more complicated in this case, comparing to what we had in
$I-IV$. The reason for it is that for $k+2={p\over q}$, $p,q$ --- integer
numbers, the bosonic screening $E$ becomes nilpotent:
\eqn\eBonil{ E^q=0}
It results essentially in that there appear more maps among the free Fock
spaces than there were for irrational central charges. Easy to see, that it
changes the diagrams only for the $(\alpha ,\beta )$ pairs where $\alpha$
satisfies the integrality condition given by \econdII, i.e., for the cases
$II,\ IV_\pm$.
Thus the diagrams of maps in Fig.1 remain the same for the cases $I,\ III$.
The proper modification of the diagram $IV$ in Fig.1 is shown in the Fig.3.
Comparing Fig.1, Fig.3. and the Felder resolution for the irreducible
"discrete" representation of the Virasoro algebra we see that passing to the
rational values of $\hat{c}$ in N=2 $SVir$ effectively amounts to using the
correct input for the Virasoro piece for the rational central charge $c_M$.

Unlike the diagrams shown in Fig.1, the diagram in Fig.3 does not admit a
natural structure of the complex. Essentially this is because of presence of
the (shaded in Fig.3) rows, corresponding to the boundary of the Kac table for
Virasoro. To obtain the complex (actually, the double complex), one throws away
these lines to come up with a picture, shown in Fig.4 --- it is a commutative
diagram and the composition of any two consequent vertical or horizontal arrows
is zero.
Now we can use a spectral sequence, similar to what we used before for the
complex in the Fig.2.
The "horizontal" cohomology again are nontrivial in only one column and give
the irreducible representations of Virasoro, via the Felder resolution.
The vertical cohomology give the resolution (here $L_{nml}$ denotes the
irreducible representation of $N=2\ SVir$, corresponding to  $(\alpha
_{nm},\beta _{nml})$, whereas $L_{nm}$ denotes the irreducible representation
of $Vir$, corresponding to $\alpha _{nm}$)
\eqn\mainres{\eqalign{
&0\rightarrow L_{nml}\rightarrow L_{nm}\otimes F(\beta _{nml})\otimes F_{gh}
\rightarrow L_{(n+1)m}\otimes F(\beta _{(n+1)ml})\otimes F_{gh}\rightarrow
\cdots  \cr
&\rightarrow L_{(q-1)m}\otimes F(\beta _{(q-1)ml})\otimes F_{gh}\rightarrow
L_{(q-1)m}\otimes F(\beta _{(q-1)m(p-m+l)})\otimes F_{gh}\rightarrow \cr
&\rightarrow L_{(q-2)m}\otimes F(\beta _{(q-2)m(p-m+l)})\otimes
F_{gh}\rightarrow \cdots \cr
&\rightarrow L_{1m}\otimes F(\beta _{1m(p-m+l)})\otimes F_{gh}\rightarrow
 L_{1m}\otimes F(\beta _{1m(l+p)})\otimes F_{gh}\rightarrow
\cdots \cr
}}
(This is $IV_-$ case, the similar resolution exists for $IV_+$.) Starting from
$L_{(q-1)m}\otimes F(\beta _{(q-1)m(p-m+l)})\otimes F_{gh}$ the resolution
becomes periodic (with the period $2(q-1)$) in its Virasoro sector. Passing to
the next period shifts the Liouville charge $\beta$ by the amount $\alpha _-p$
(cf. the charges of the first and the last a Fock spaces in \mainres). Note
also that in \mainres participate only the representations $L_{nm}$ from the
principal Kac table (i.e. with $0<n<q,\ 0<m<p$).

 The "throwing away" procedure that we use seems a little bit {\it ad hoc}. In
fact, it can be understood using the same logic as we used in the previous
section for  irrational $\hat{c}$. Another way to obtain the resolution
\mainres is explained in the next section.

Let us compute the characters of the representations of the type
$IV$ here, using the resolution \mainres.
The formulas are (unlike the previous subsection, these are just the usual
nonnormalized characters):
\eqn\ercharmn{\eqalign{
&\chi _{nml}=x^{-2\alpha _-\beta _{nml}}\chi _{bc}(t,x)
\sum_{r=1}^{q-1}(-1)^{r-n}x^{r-n}
\chi ^{Vir}_{rm}(t)\biggl( \sum_{\scriptstyle s=0\ if\ r\geq n \atop
			       \scriptstyle s=1\ if\ r<n}^\infty x^{2sq}t^{\Delta ^+(s)} \cr
& +x^{2(q-r)} \sum_{s=1}^\infty x^{2sq}t^{\Delta ^-(s)}\biggr) \cr
&\Delta ^+(s)=-\beta _{rm(l+sp)}(\beta _{rm(l+sp)}-\alpha _++\alpha _-) \cr
&\Delta ^-(s)=-\beta _{rm(l-m+sp)}(\beta _{rm(l-m+sp)}-\alpha _++\alpha _-) \cr
}}
--- for the $IV_-$ case and
\eqn\ercharpl{\eqalign{
&\chi _{nml}=x^{-2\alpha _-\beta _{nml}}\chi _{bc}(t,x)
\sum_{r=1}^{q-1}(-1)^{r-n}x^{r-n}
\chi ^{Vir}_{rm}(t)\biggl( \sum_{\scriptstyle s=0\ if\ r\leq n \atop
		            \scriptstyle s=1\ if\ r>n}^\infty
x^{2sq}t^{\Delta ^+(s)} \cr
& +x^{2(q-r)} \sum_{s=0}^\infty x^{2sq}t^{\Delta ^-(s)}\biggr) \cr
&\Delta ^+(s)=-\beta _{rm(l+sp)}(\beta _{rm(l+sp)}-\alpha _++\alpha _-) \cr
&\Delta ^-(s)=-\beta _{rm(l+m+sp)}(\beta _{rm(l+m+sp)}-\alpha _++\alpha _-) \cr
}}
--- for the $IV_+$ case. Because the symbol $q$ is reserved already for the
index of the minimal model, we denoted the modular parameter by $t$. $\chi
^{Vir}_{rm}(t)$ and $\chi _{bc}(t,x)$ stand for the charaters of the Virasoro
irreducible representation $L_{rm}$ the $(b,c)$ ghosts Fock space (the latter
is independent of $n,m,l$) respectively.

It should be noted that the formulas for the characters were also obtained by
Kac and Wakimoto and, independently, in \Chung. Both groups used the different
methods, which did nott allow to see the role of the Virasoro characters.

It is particularly interesting to consider the representations $IV_+$ with
$l=jp$ and $IV_-$ with $l=-1+jp$, where $j$ is any integer number.
For definiteness let us consider the latter.

Note, that for such $l$ the charges $\beta$ of the Liouville Fock spaces
coupled to the Virasoro irreducible representation $L_{mn}(Vir)$ in the
resolution \mainres are exactly the same as in the Lian-Zuckerman's papers
\ref\LZI{B.~H.~Lian, G.~J.~Zuckerman {\it Phys.~Lett {\bf 254B}(1991)
417,}}\ref\LZII{B.~H.~Lian, G.~J.~Zuckerman {\it Phys.~Lett {\bf 266B}(1991)
21,}}\ref\LZIII{B.~H.~Lian, G.~J.~Zuckerman {\it Comm.~Math.~Phys.~{\bf
145}(1992) 54}} on the spectrum of 2-d gravity; it means that these liouvilles
are the "dressing fields" of the discrete states. More precisely, the weights
$\Delta^\pm(s)$ in \ercharmn are related to the weights of the  singular
vectors  in {\it the Verma module}
$M_{mn}(Vir)$ (whose quotient is $L_{mn}(Vir)$) by the formula:
\eqn\erelng{
\Delta ^\pm(s)+\Delta _{nm}(-j-s)=1
}
In this formula, $\Delta _{nm}(i)$ denotes the weight of the singular vector in
$M_{mn}(Vir)$, and $i$ is the ghost number of the LZ state, corresponding to
this singular vector.

It is interesting, that the the infinite number of LZ states (with the ghost
numbers $\leq -j$) are combined in the single object --- the character of the
irreducible representation of $N=2$ SuperVirasoro algebra. It demonstrates the
relationship between 2-d gravity and $N=2\ SVir$ very explicitly at the level
of representations. In Section 4.2 we shall learn more about this.

\newsec{Details of computations and proof of the equivalence theorem}
\subsec{Reduction of the BRST complex }

Now we can compute the cohomology $H^*_{Q_R}(L_k\otimes Wak_{-k-4})$.
It is convenient to take the Felder resolution of the irreducible
representation $L_k$ of $\widehat{sl}(2)$ in terms of free Fock modules
generated by the currents $\partial x(z),\ \beta _M(z),\ \gamma _M(z)$\foot{
As the Toda-Liouville sector is already bosonized, we use the subscript "T" for
its $\beta ,\gamma$ system, so this sector is generated by the currents
$\partial \varphi (z), \beta _T(z),\gamma _T(z)$.},
 using a usual screening operator
\eqn\esuscrI{E^{(-)}_{sl(2)}=\oint :\beta _M(z)e^{\sqrt{2}\alpha _-x(z)}:
}
 In principle, there is another screening
\eqn\esuscrII{E^{(+)}_{sl(2)}=\oint :(\beta _M(z))^{-(k+2)}e^{\sqrt{2}\alpha
_+x(z)}:
}
Although the notion of the rational powers of $\beta _M(z)$ can be justified if
we make bosonisation of the $\beta _M, \gamma _M$ pair, we prefer to deal with
the "conventional" choice \esuscrI.

We know that under the standard (Drinfeld-Sokolov) hamiltonian reduction these
two screenings go to the two screenings
\eqn\eViscr{E^{(\pm)}_{Vir}=\oint :e^{\alpha _\pm x(z)}:
}
of the Virasoro algebra.
Thus we can anticipate that it is also so in {\it our} reduction scheme, which
means that the "conventional" screening $E^{(-)}_{sl(2)}$ must go to
$E^{(-)}_{Vir}$ and the "unconventional" one  $E^{(+)}_{sl(2)}$ must go to
$E^{(+)}_{Vir}$.

Let us decompose the reduction BRST operator
$Q_R$
as
\eqn\estddec{\eqalign{
&Q_R=\hat{Q}_R+c^0(0){\cal H}_0 \cr
&{\cal H}_0=(H_1+H_2+2:b_+c^+:-2:b_-c^-:)_0 \cr
}}

and note that there is a relation
\eqn\etriv{ {\cal H}_0=\{Q_R,b_0(0)\}}

Then the usual argument shows that on the cohomology
\eqn\etriv{ {\cal H}_0|_{H^*_{Q_R}}=0}
must hold.
Moreover, we see that the operator $c^0(0)$ is $Q_R$-nontrivial so it maps  the
$Q_R$ cohomology to itself:
\eqn\ecmap{
|\lambda >\in H^*_{Q_R}\longrightarrow c^0(0)|\lambda >\in H^*_{Q_R}}
(This "doubling" of $Q_R$ cohomology will be important in relations to 2-d
gravity.)

It is convenient to choose the "light-cone" coordinates in the $\beta ,\gamma$
sector:
\eqn\elcbg{\eqalign{
&\beta _+={1 \over \sqrt{2}}(\beta _M+\beta _T) \cr
&\beta _-={1 \over \sqrt{2}}(\beta _M-\beta _T) \cr
&\gamma _+={1 \over \sqrt{2}}(\gamma _M+\gamma _T) \cr
&\gamma _-={1 \over \sqrt{2}}(\gamma _M-\gamma _T) \cr
}}
To compute $H^*_{Q_R}$  let us use a double complex with differentials
\eqn\edfrnt{\eqalign{
&d_1=\sqrt{2}\oint \big(c^+\beta _+ -\sqrt{2}c^0(:\beta _+\gamma _+:-:b_+c^+:)
\big) \cr
&d_2=\oint c^0(-\sqrt{2}\alpha _+(\partial \varphi +\partial x) -2:\beta
_-\gamma _-: -2:b_-c^-:) \cr
}}
Thus we have managed to completely separate in $d_1$ and $d_2$ two subsets of
fields: $\{c^+(z),b_-(z),\beta _+(z),\gamma _+(z)\}$ and
$\{c^0(z),b_0(z),c^-(z),b_-(z),x(z),\varphi (z),\beta _-(z)$, \break $\gamma
_-(z)\}$.
It means that our double complex is in fact a direct product of two complexes
with differentials respectively $d_1$ and $d_2$. Their cohomology can be
computed independently of each other.
Let us  compute the cohomology of $d_1$ first.
It is easy to see that in $H^*_{d_1}$ two systems $\beta _+, \gamma _+$ and
$b_+,c^+$ "cancel" each other. It means that there are no excitations along
these four directions in the "physical" space (i.e. $H^*_{d_1}$). This is an
example of the famous Kugo-Ojima quartet decoupling mechanism.
To compute  $H^*_{d_2}$ we need to bosonise the $\beta _-, \gamma _-$ system in
terms of two free bosons $\psi (z),\ \chi (z)$:
\eqn\ebosbega{\eqalign{
&:\beta _-\gamma _-:(z)=\partial \psi (z) \cr
&\beta _-(z)=e^{(\psi +i\chi )} \cr
}}
 To be precise, following \ref\FeFrII{ B.~Feigin, E.~Frenkel
{\it Comm.~Math.~Phys.~{\bf 128}(1990) 161}}, the Hilbert space $F_{\beta ,
\gamma}$   of the $\beta _-, \gamma _-$ system is represented by the (0-)
cohomology of the screening operator
\eqn\ebgscr{
F=\oint :exp(-i\chi (z)):
}
acting on the free Fock modules
\eqn\eFres{
\bigoplus_{...}F_{}(\psi )\otimes F_{}(\chi )}
(so it is a free field resolution).
Now we have a double complex again: one differential is $d_2$ which acts now by
the "free field" formula
\eqn\ebdII{
d_2=\oint \big(-\sqrt{2}\alpha _+(\partial \varphi (z)+\partial x(z))
-2\partial \psi (z):-2:b_-c^-:\big)
}
 and the other one is $F$.
To compute the cohomology of this double complex let us find $H^*_{d_2}$ first.
It is easy to do because it is actually the $U(1)$ cohomology.
The representatives of $H^*_{d_2}$ can be written as:
\eqn\erepI{\eqalign{
&X(z)=\sqrt{2}x(z)-\alpha _+(\psi +i\chi ) \cr
&\phi (z)=\sqrt{2}\varphi (z)+\alpha _+(\psi +i\chi ) \cr
&B(z)=b_-e^{(\psi +i\chi )}(z) \cr
&C(z)=c^-e^{-(\psi +i\chi )}(z) \cr
}}
Together they form a space
\eqn\erepspaI{
\bigoplus F_{}(X)\otimes F_{}(\phi )\otimes F(B,C)
}
Now we should compute $H^*_F(H^*_{d_2})$ (the second term of the spectral
sequence of the double complex).
If we only need the $Q_R$ cohomology of two Wakimoto modules, we are done, and
the answer is represented by the free field resolution
\eqn\intans{
F_{\alpha \beta}\rightarrow F_{\alpha -{\alpha _+\over 2}\beta -{\alpha _+\over
2}}\rightarrow F_{\alpha -\alpha _+ \beta -\alpha _+}\rightarrow \cdots
}
The charges $\alpha ,\beta$ in terms of the weights $J_M$, $J_T$ of the
Wakimoto modules are given by $\alpha =\alpha _-J_M$, $\beta =-\alpha _-J_T$.
The differential in \intans is $F$.

 However, if we are interested in the $Q_R$ cohomology of the product of the
irreducible representation times Wakimoto module, we should remember about the
screening
$E^{(-)}_{sl(2)}$ and the Felder resolution we made.
Note, that the $\widehat{sl}(2)$ screenings $E^{(\pm )}_{sl(2)}$ can be
represented by $E^{(\pm)}_{Vir}=\oint :e^{\alpha _\pm X(z)}:$ just as we
thought (cf. \esuscrI,\esuscrII):
\eqn\escrIII{\eqalign{
&:e^{\alpha _+ X(z)}:=:e^{\sqrt{2}\alpha _+ x(z)-(k+2)(\psi +i\chi )}:=(\beta
_M)^{-(k+2)}:e^{\sqrt{2(k+2)}x(z)}: \cr
&:e^{\alpha _- X(z)}:=:e^{\sqrt{2}\alpha _- x(z)+(\psi +i\chi )}:=\beta
_M:e^{-{2\over \sqrt{2(k+2)}}x(z)}: \cr
}}
(By the triviality of  $H^*_{d_1}$ we may simply set $\beta _-(z)=\beta _M(z)$
at the level of the chiral algebra. In \escrIII we use this relation.)

 Also, adding the trivial piece
\eqn\etrip{
\{d_2,b_0(z)\}=-\sqrt{2}\alpha _+(\partial \varphi (z)+\partial x(z))
-2\partial \psi (z):-2:b_-c^-:}
to  $i\chi (z)$ and using \ebosbega, \erepI  we see that $F$ \ebgscr can be
represented as
\eqn\enewF{
F=\oint :exp(-i\chi (z)):=\oint B(z):e^{-{\alpha _+ \over 2}(X(z)+\phi (z))}
}
so actually there is no abuse of notations in \ebgscr and we may think of $F$
just
as of the fermionic screening from the representation theory of N=2 $SVir$
which we introduced in Sec 3.

Thus, finally, we are left with the cohomology of the double complex (the last
one in this story) with two differentials. One of them is just $F$ and another
one comes from the differential of the Felder resolution and is given by the
certain powers of the screening $E^{-}$.
This "ultimate" double complex is shown in the Fig.5.

It is interesting to compare this complex with the double complex in the Fig.4
because we anticipate that their cohomology are the same (and give the
irreducible representation of N=2 Supervirasoro algebra). Although these two
look very similar, there are some differences. The basic distinction between
them is that the "horizontal" differentials are "made" of two different
$Vir$-screenings: $E^{(+)}$ for the complex in the Fig.4 and $E^{(-)}$ for that
one in the Fig.5.

They  represent two  different choices of the Felder resolutions for the same
irreducible representation of Virasoro. As $E^{(-)}$ and $F$ commute (cf.
\ecomm), the vertical differential in Fig.5 is always $F$. The relations
between $E^{(+)}$ and $F$ are more involved. As a result in the Fig.4 the
vertical differential is either $F$ (denoted by $F_1$) or a combination like
$xE^{(+)}F+FE^{(+)}$ (denoted by $F_2$), or $FE^{(+)}F$, depending on the place
in the complex. Then, the weights $\alpha$ corresponding to the principal
Virasoro Kac table representations are concentrated along one column in Fig.4.
There are no weights from the boundary of the Kac table (we "threw them away").
On the other hand, the boundary weights are present in the Fig.5 and the
weights from the interior of the principal Kac table are situated along the
"shifted" vertical segments, shaded in the picture. (A shift occurs each time
we pass through the horizontal line corresponding to the boundary weight.)

These differences result in the difference in computation of the cohomology.
In both cases one can use the "vertical" filtration of the double complex to
compute the "horizontal" cohomology first. In the case of the complex in the
Fig.5 we immediately had the Kac table irreducible representations of Virasoro
times some Fock spaces of ghosts and Liouville field all along one column. The
"vertical" differential was induced either from $F$ or $xE^{(+)}F+FE^{(+)}$ or
$FE^{(+)}F$. The latter combination appeared between the rows where we have
thrown away the "boundary  row".

On the other hand, in the case shown in the Fig.6, the horizontal cohomology
give the irreducible representations for the rows corresponding to the weights
from the interior of the principal Kac table and zeros for the rows
corresponding to the weights from the boundary of the Kac table.  We obtain
zero cohomology for the "boundary" rows because each such row represents two
(left- and right-sided) glued together resolutions of the same irreducible
representation, so the resulting horizontal complex is exact.  The nontrivial
cohomology are concentrated along the "shifted segments".

The "vertical" cohomology gives again the resolution \mainres. To see this we
should first recall that the three pairs of indeces $(n,m)$, $(n+q,m+p)$ and
$(q-n,p-m)$ describe the same Virasoro weight. Note then that the "knight move"
differential
$d_2$ is nontrivial for this  spectral sequence. It is a "connecting
differential" --- it acts from the top of one "segment" to the bottom of
another "shifted segment", mending all these segments together in one complex
which is just \mainres.

Now to get $H^*_{Q_R}(L_k\otimes Wak_{-k-4})$ we recall about the $c^0$ zero
mode "doubling". (Here also, as in the case of  $H^*_{Q_R}(Wak_k\otimes
Wak_{-k-4}$ there could be a problem with the "knight move" differential of the
main spectral sequence, but it is zero here for the same reason as there.)
Thus we have
\eqn\eFnl{
H^*_{Q_R}(L_k\otimes Wak_{-k-4})=\bigl[L(N=2\ SVir)\oplus c^0_0 L(N=2\ SVir)
\bigr]
}

Let the $sl(2)$ weights (spins) of the modules $L_k$ (the matter) and
$Wak_{-k-4}$ (the "Toda-Liouville") in \eFnl be respectively $J_M$ and $J_T$.
Then the parameters $(\alpha ,\beta )$ of the irreducible representation
$L_{\alpha ,\beta}(N=2\ SVir)$ there are given by
\eqn\eweig{\eqalign{
&\alpha =-\alpha _-J_M \cr
&\beta  =\alpha _-J_T \cr
}}
(remember that $\alpha _-=-{1 \over \sqrt{k+2}}$)

In particular, let $L_k(J_M)$ be the admissible irreducible representation of
$\widehat{sl}(2)$, i.e.
\eqn\mwei{
J_M={m-1 \over 2}+(k+2){1-n \over 2}}
and let the weight $J_T$ of $Wak_{-4-k}(J_T)$ be such that it dresses properly
a singular vector (having weight $J'_M$) in the Verma module $M_k(J_M)$, which
means that $J_T+J'_M=-1$. Then the representation $L_{\alpha \beta}(N=2\ SVir)$
in \eFnl in the notations of Section 3
is $L_{nml}$ of $IV_-$ type with $l=-jp$, where $j$ is the ghost number of the
corresponding $\widehat{sl}(2)$-BRST state.

Now, to prove \ecca (it was promised in the Sec.2), we take the vacuum
representation of $\widehat{sl}(2)$ which is a VOA for $\widehat{sl}(2)$ and,
looking at \eFnl, recall that the vacuum representation of N=2 $SVir$ algebra
we see in the right hand side is a VOA for the latter. (In the more physical
language it just means that by the $Operators\rightarrow States$ correspondence
the chiral algebras $N=2\ SVir$ and $\widehat{sl}(2)$ are related to the
"descendants of the unity operator" which form the corresponding vacuum
representations.)

\subsec{Equivalence of the spectra of coset and 2-d gravity.}

Finally, we compute the cohomology $H^*_{Q_W}$ of the reduced $\widehat{sl}(2)$
--- BRST complex $H^*_{Q_R}(L_k\otimes Wak_{-k-4})$ and show they coincide with
the 2-d gravity BRST cohomology of $L_k^{DS}\otimes Wak_{-k-4}^{DS}$\foot{
As above, the superscript "DS" denotes the standard (Drinfeld-Sokolov)
reduction.}. For definiteness, we do it for this choice of $\widehat{sl}(2)$
representations, but it can easily be done also for other choices, mentioned in
Section 2.

To do the actual computation, let us use the equivalence of two complexes:
\eqn\esuredII{(H_{Q_R}^*(L_k(\widehat{sl_k}(2))\otimes
Wak_{-4-k}),{Q_W})=(L(N=2\ SVir)\oplus c^0_0L(N=2\ SVir),G_0^+)
}
Suppose that the spin of $L_k(\widehat{sl_k}(2))$ is given by \mwei and the
spin of the Wakimoto module $Wak_{-4-k}$ is "dressing" (see the discussion
after \mwei). Then the $N=2$ irreducible representation $L(N=2\ SVir)$ in
\esuredII is $L_{nml}$ with $l=-jp$.

Now, take the "2-d gravity" resolution \mainres of $L_{nml}$. On the "2-d
gravity" modules
$L(Vir)\otimes F(Liouv.)\otimes F_{gh}$ the action of the differential $G_0^+$
is given by the formula \ediff: $Q_W=G^+_0=Q_{Vir}$.
So we have a double complex. One  differential there is $Q_{Vir}$ and the other
one comes from the resolution \mainres.

The cohomology of this double complex, by our construction, compute the
$\widehat{sl}(2)$ BRST cohomology of $L_k(\widehat{sl_k}(2)\otimes Wak_{-4-k}$
--- the physical states of the coset model. At this stage we have to assume
that {\it we know} all the physical states {\it either} for the coset model
{\it or} for 2-d gravity. Then we shall be able to find the the spectrum of
states for the other theory. Suppose for definiteness that we know the spectrum
of 2-d gravity coupled to $(q,p)$ -- minimal matter\foot{
Using the standard homological algebra, it is easy to go  the other way ---
i.e. to obtain the 2-d gravity spectrum {\it from} the spectrum of the coset.}

It means that in our double complex we can compute the "gravitational"
cohomology $H^*_{Q_{Vir}}$ first --- it gives the first term of the spectral
sequence. Recalling that in \mainres for the representations $IV_-$ with
$l=-jp$ the charges of the liouvilles are just right to dress  the "discrete
states", we end up with the situation shown in the Fig.6. It is important to
remember that we compute the usual, i.e. "absolute" cohomology. It is well
known that the BRST cohomology
of the properly dressed irreducible representation are given by {\it two}
elements at the adjacent ghost numbers. This "Virasoro doubling" is due to the
zero mode of the diffeomorphisms ghost $C_0$. Thus
all nontrivial cohomology states in the first term of the spectral sequence
(the Lian-Zuckerman states) are concentrated along the "shifted segments"
according to their ghost numbers. Each time we pass half-period $p-1$ (recall
the periodic structure of \mainres), we get a shift by -1 of the ghost number.
At the ghost number zero there are two states, and at each positive ghost
number, there are four states. Note the "$\widehat{sl}(2)$ doubling" due to the
zero mode $c_0^0$.

To compute the second term of the spectral sequence, we need the properties of
the differential $d$ of the resolution \mainres. This operator was studied in
\rDots. Remember that it is induced either from the fermionic screening $F$
(when it acts "within one half-period") or from $FEF$ (when it acts "between
two half-periods"). Thus it changes the ghost number by -1 or by -2 units
respectively. Its action on the LZ states is shown by the arrows. We see that
the cohomology are concentrated in the same two degrees as the cohomology
$H^*_{Q_{Vir}}(L_{nm}\otimes F(\beta _{nml}))$ and also have two elements, now
due to the "$\widehat{sl}(2)$ doubling". Of course, this is the correct answer
for $H^*_{Q_{BRST}}(L_k\otimes Wak_{-k-4})$. The nontrivial thing that happens
is that the "Virasoro doubling" gets transformed into the
"$\widehat{sl}(2)$ doubling".

Now we should only note that $L_{nm}\otimes F(\beta _{nml})$ --- the first term
of the resolution \mainres, --- is nothing else but the Drinfeld-Sokolov
reduced
$(L_k)^{DS}\otimes (Wak_{-k-4})^{DS}$. Thus we have shown, that the coset- and
2-d gravity cohomology are the same thing, basically because they both describe
the $G_0^+$ cohomology of the $N=2\ SVir$ irreducible representation $L_{nml}$.

\bigbreak\bigskip\bigskip\centerline{{\bf Acknowledgements}}\nobreak
I am gratefull to M.~Bershadsky, E.~Frenkel, C.~Vafa for the interesting
discussions. I thank S.~Chung for bringing to my attention the paper \Chung.

Research supported in part by the Packard Foundation and by NSF grant
PHY-87-14654

\footatend\vfill\supereject\immediate\closeout\rfile\writestoppt
\baselineskip=14pt\centerline{{\bf References}}\bigskip{\frenchspacing%
\parindent=20pt\escapechar=` \input refs.tmp\vfill\eject}\nonfrenchspacing
\bye